\documentclass[aps,prl,amssymb,groupedaddress,nofootinbib]{revtex4}

\usepackage[english]{babel}
\usepackage{amsfonts}
\usepackage{amssymb}
\usepackage{epsfig}
\usepackage{amsmath}
\usepackage{fancyhdr}
\usepackage{textcomp}
\usepackage{setspace}

\def\L{\Lambda}

\begin{document}

\title{Minimal cut-off vacuum state constraints from CMB bispectrum statistics}

\author{P. Daniel~Meerburg$^{1}$}

\author{Jan Pieter~van der Schaar$^{2}$}

\affiliation{$^1$ Astronomical Institute ``Anton Pannekoek", University of Amsterdam,\\
Science Park 904, 1098XH Amsterdam, The Netherlands\\
$^2$  Korteweg-de Vries Institute for Mathematics, University of Amsterdam,\\ 
Plantage Muidergracht 24, 1018 TV Amsterdam, The Netherlands}

\date{\today}

\begin{abstract}

\noindent 
In this short note we translate the best available observational bounds on the CMB bispectrum amplitudes into constraints on a 
specific scale-invariant New Physics Hypersurface (NPH) model of vacuum state modifications, as first proposed by Danielsson, 
in general models of single-field inflation. As compared to the power
spectrum the bispectrum constraints are less ambiguous and provide an 
interesting upper bound on the cut-off scale in general models of 
single-field inflation with a small speed of sound. This upper bound
is incompatible with the power spectrum constraint for most of the 
parameter domain, leaving very little room for minimal cut-off vacuum state 
modifications in general single-field models with a small speed of sound. 
\end{abstract}

\pacs{1}

\maketitle


\section{Introduction}

Bounds on higher-order non-Gaussian contributions to the observed and
approximately Gaussian temperature anisotropy map of the 
Cosmic Microwave Background are rapidly entering a phase where they
can be used to constrain and sometimes even rule out certain (exotic) 
models of inflation. For instance, bounds on the amplitude of the CMB 
three-point function, or bispectrum, have already narrowed down the parameter 
range of string inspired Dirac-Born Infeld models of inflation. 

In a separate development it was recently shown that modifications to the inflationary vacuum 
state can also result in significantly enhanced non-Gaussian signals
\cite{Chen2007,HolTol:2007, Meerburg2009a,Meerburg2009b}. In particular the bispectrum can reveal enhanced features that
essentially arise due to interactions between excited quanta in the modified vacuum. 
The power spectrum is also sensitive to deviations from the
Bunch-Davies vacuum, but there the corrections are suppressed
by the Bogoliubov parameter, whose magnitude is small and typically 
related to the inflationary Hubble parameter divided by the scale of
new physics, which is already constrained to be smaller than $10^{-2}$
\cite{Pspectrum-constraints}.  For the bispectrum one instead encounters
an additional enhancement factor, independent of the Bogoliubov parameter, that is in 
fact proportional to the scale where effective field theory is
supposed to break down. The appearance of this enhancement factor can be traced back 
to the New Physics Hypersurface (NPH) used to define the modified
vacuum state. The higher the cut-off scale representing new physics, the more 
time the fields have before they cross the horizon and freeze, and the
more time the interactions effectively have to generate a significant 
non-Gaussian contribution. This also implies that the (decoupling)
limit of taking the cut-off scale to infinity in a modified vacuum is
in general ill-defined, producing an infinite bispectrum, 
whereas the power spectrum yields the standard result in 
this decoupling limit (assuming the Bogoliubov parameter 
is suppressed by one over the cut-off scale).  

In this work the most recent bispectrum constraints will be applied to
a particular and arguably the simplest model of vacuum state
corrections, which was first considered in \cite{EGKS}, but
constructed and applied more generally by Danielsson \cite{NPH}. 
We will derive to what extent this vacuum state model 
can be ruled out by the bispectrum data alone in two particular
scenarios; a slow-roll single field inflationary model with a higher derivative operator 
representing the interactions and a non-canonical single field model
with a small speed of sound. The bispectrum prediction for a general (NPH) vacuum 
state modification in these single-field models was computed in
\cite{Meerburg2009a, Meerburg2009b}. As an important and hopefully
illuminating illustration we will slightly extend and improve the
analysis, and apply it to the specific vacuum state proposal of Danielsson. 

This paper is organized as follows. We will start with a quick review
and small generalization of the Danielsson vacuum state proposal, followed by 
a summary of the results for the bispectrum in the context of a vacuum
state modification in general single-field models of inflation. We then explain
how these results can be turned into an {\it upper bound} on the scale
of new physics and to what extent this vacuum state proposal is ruled out or not. 
We end with some conclusions and prospects for future improvement of the bounds. 

\section{A minimal model of vacuum state modifications}

Let us begin with a short review of Danielsson's original proposal for a modified vacuum state, under the natural assumption that a high energy 
cut-off scale $\L_c$ exists beyond which effective field theory breaks down \cite{NPH}. Solving for the mode functions in an inflating (pure de Sitter) background
one writes down solutions for the field and conjugate momentum operators in terms of creation and annihilation operators in the standard way. 
Now identify a (conformal) time $\eta_0$ serving as the initial time where one would like to define the vacuum state. This initial time $\eta_0$ is related 
to the cut-off $\Lambda_c$ by demanding that the physical momentum at $\eta_0$ equals the cut-off scale, i.e. $|k \eta_0|=\frac{\Lambda_c}{H}$. 
This ensures that the description is always within the effective field theory regime. Note that since we are imposing a cut-off on the physical momentum 
the initial time $\eta_0$ is necessarily a function of the comoving momentum $k$. With respect to this initial time $\eta_0$ 
the creation operators (and subsequently the annihilation operators) can then be expressed as 
\begin{equation}
a_k(\eta) = u_k(\eta) \, a_k(\eta_0) + v_k(\eta) a_{-k}^\dagger(\eta_0) \, ,
\label{aauv}
\end{equation}
describing nothing else but the mixing of creation and annihilation operators as time progresses. 
To ensure the time-independent commutation relations for the creation- and annihilation operators requires that  $|u_k|^2 - |v_k|^2=1$.
Let us now define a natural candidate for a vacuum state at some initial conformal time $\eta_0$
\begin{equation}
a_k(\eta_0) \left| 0, \eta_0 \right> = 0 \, .
\end{equation}
Obviously this choice requires that $v_k(\eta)$ vanishes at $\eta=\eta_0$ and as a direct consequence this candidate vacuum corresponds 
to a minimal uncertainty state at $\eta=\eta_0$ \cite{NPH}. This choice of vacuum can be understood as selecting the `local' empty state 
at the time the physical momentum equals the high-energy cut-off $\Lambda_c$. This time will be different for different comoving momenta
and defines a so-called `New Physics Hypersurface'. It can therefore be viewed as the vacuum state that is closest to the Bunch-Davies state 
in the presence of a high-energy cut-off and can in that sense be regarded as a minimal modification. 

To understand the relation of this choice of vacuum to the standard Bunch-Davies state, consider
the field operator which should be proportional to the sum of the creation and annihilation operators. In terms of mode-function solutions to the equations 
of motion $f_k(\eta)$ this implies that  
\begin{equation}
f_k(\eta) = N_k \left( u_k(\eta) + v_k^\star (\eta) \right) \, ,
\end{equation}
where the overall (real) normalization $N_k$ is fixed by the Klein-Gordon normalization condition. 
Similarly, the expression for the canonical momentum operator defines
a function $g_k(\eta)$, which is proportional to the difference between $u_k(\eta)$ and $v_k(\eta)$. 
\begin{equation}
g_k(\eta) = \tilde{N}_k \left( u_k(\eta) - v_k^\star (\eta) \right) \, .
\end{equation}
Using these relations one can derive the following expression for the function $v_k(\eta)$ in terms of $f_k$ and $g_k$
\begin{equation}
v_k^{\star}(\eta) = \frac{1}{2} \left[ N_k^{-1} \, f_k(\eta) - \tilde{N}_k^{-1} \, g_k(\eta) \right] \, .
\label{vkeqn}
\end{equation}
Defining a vacuum state $a_k(\eta_0) \left| 0, \eta_0 \right> = 0$ is now explicitly seen to be equivalent to picking a mode-function solution $f_k$ (and its 
canonically conjugate function $g_k$) such that $v_k(\eta)$ vanishes at some $\eta=\eta_0$. Bunch-Davies mode-functions have the property that $v_k(\eta)$
only vanishes in the limit $\eta \rightarrow -\infty$, i.e. the Bunch-Davies state in this class of vacua corresponds to the minimal uncertainty vacuum
state in the infinite past. 

Now let us instead consider the general solution, which can be constructed from the Bunch-Davies mode-function and its complex conjugate
\begin{equation}
f_k (\eta) = A_k \, f_k^{BD}(\eta) + B_k \, {f_k^{BD}}^{\star}(\eta) \, ,
\end{equation}
where $A_k$ and $B_k$ are complex coefficients satisfying $|A_k|^2 - |B_k|^2=1$. 
Similarly, the momentum mode-function $g_k(\eta)$ can be expressed in terms of the Bunch-Davies momentum mode-functions
\begin{equation}
g_k (\eta) = A_k \, g_k^{BD}(\eta) - B_k \, {g_k^{BD}}^{\star}(\eta) \, .
\end{equation}
To find the natural candidate vacua at some fixed time $\eta_0$ one demands that $v_k(\eta_0)=0$. Using (\ref{vkeqn}) this gives the following expression for 
the Bogoliubov rotation parameter $\beta_k \equiv \frac{B_k}{A_k}$ in terms of the Bunch-Davies field and momentum mode-functions
\begin{equation}
\beta_k = {{\frac{N_k}{\tilde{N}_k} g_k^{BD}(\eta_0) - f_k^{BD}(\eta_0)} \over {\frac{N_k}{\tilde{N}_k} {g_k^{BD}}^{\star}(\eta_0) + {f_k^{BD}}^{\star}(\eta_0)}} \, .
\label{betamodefn}
\end{equation} 
In terms of the (Bunch-Davies) functions ${u_k}^{BD}(\eta)$ and ${v_k}^{BD}(\eta)$ that appear in the expression for the annihilation and creation operators 
(\ref{aauv}) the result simply reads
\begin{equation}
\beta_k = - {{{v_k^{\star}}^{BD}(\eta_0)} \over {{u_k^{\star}}^{BD}(\eta_0)}} \, .
\label{betauv}
\end{equation}  
A couple of comments are in order. Since the (late time) power spectrum of fluctuations
is proportional to $|f_k|^2$, the Bogoliubov rotation parameter $\beta_k$ denotes the leading order correction to the power spectrum due to  
a vacuum state modification.Scale invariant vacuum states of course require that $\beta_k$ is independent of the comoving momentum $k$, 
which will be guaranteed by the relation $k \eta_0 = -\frac{\Lambda_c}{H}$, i.e. $\eta_0$ is $k$-dependent.  
Also note that this minimal cut-off vacuum construction has so far been completely general, not depending on any details of the 
inflationary Lagrangian. Whatever inflationary model one is interested in, to construct these states and determine the Bogoliubov parameter $\beta_k$ 
one simply plugs in the relevant (mode-) functions in the inflationary case of interest.

The best known (and originally discussed) example is that of the massless scalar field, that is associated to standard slow-roll inflation. 
The Bunch-Davies field and momentum mode-function in that case read
\begin{equation}
f_k^{BD}(\eta) = \frac{1}{\sqrt{2k}} e^{-ik\eta} \left( 1- \frac{i}{k\eta} \right) \quad , \quad g_k^{BD}(\eta) = \sqrt{\frac{k}{2}} e^{-ik\eta} \, .
\end{equation}
Using the expression (\ref{betamodefn}) or (\ref{betauv}) one then derives the following expression for the Bogoliubov parameter $\beta_k$
\begin{equation}
\beta_k = \frac{i}{2k\eta_0 + i} e^{-2i k\eta_0} \, .
\label{eq:bogslowroll}
\end{equation}
Since $|k \eta_0| = \frac{\Lambda_c}{H} \gg 1$ this Bogoliubov parameter is indeed scale invariant and approximates to  
$\beta_k \approx \frac{H}{2\Lambda_c} e^{i (\frac{3}{2}\pi - \frac{2\Lambda}{H})}$.

As we have emphasized, the construction of these minimal cut-off states is completely general and can just as well be applied in the context of
small speed of sound models of inflation, which includes DBI inflation
as a special class. One might suggest that in non-canonical
  models of inflation the introduction of a cut-off scale is not
  required, since an infinite number of higher order corrections have
  been resummed leading to the non-canonical kinetic
  term. However, these corrections are only a subset of all the
  possible higher-dimensional operators that are expected to
  contribute as the string- (or cut-off) scale is approached,
  motivating the general introduction of a cut-off scale\footnote{More
    specifically, in \cite{Kinney2007} it was shown that in DBI models of
    inflation under certain conditions the standard procedure for defining the Bunch-Davies
    vacuum can fail.}.
So the main, and for our purposes only, difference between 
small speed of sound and slow-roll models of inflation is the introduction of a reduced (and assumed constant) speed of sound $c_s < 1$ of the 
inflaton fluctuations. For the Bunch-Davies functions $u_k^{BD}$ and $v_k^{BD}$ one can check that the only effect is the appearance of a single factor 
of the speed of sound $c_s$ in the $k\eta_0$ terms as compared to the standard slow-roll case. The end-result for the Bogoliubov parameter therefore reads
\begin{equation}
\beta_k = \frac{i}{2k c_s \eta_0 + i} e^{-2i k c_s \eta_0} \, ,
\label{eq:bogsmallcs}
\end{equation}
which translates into the scale-invariant approximate expression $\beta_k \approx \frac{H}{2 c_s \Lambda_c} e^{i (\frac{3}{2}\pi - \frac{2 c_s \Lambda}{H})}$. 
One observes that the ratio $H/\Lambda_c$ governing the magnitude of the corrections in standard slow-roll is modified to $H / c_s \Lambda_c$
in these models, which could be orders of magnitude bigger. Note that the length scale $1/H^* \equiv c_s/H < 1/H$ corresponds to the sound horizon in 
small speed of sound models, corresponding to the length scale at
which the behavior of the mode-functions turns non-adiabatic. The
appearance of the factor  
$k c_s \eta_0$ is a general feature in these models, which was for instance also observed for the enhancement factors in the three-point function 
that appear whenever one introduces an arbitrary modified initial state at some high-energy cut-off. We conclude that the same
is true for the Bogoliubov parameter in cut-off modified initial state proposals: the magnitude (and phase) depends on the combination $k c_s \eta_0$ 
which results in larger absolute values of the Bogoliubov parameter for the same value of the cut-off and inflationary Hubble parameter, as compared
to standard slow-roll models. 

This ends our short summary of minimal cut-off initial states. The results for the three-point function, as reported in \cite{Meerburg2009a, 
Meerburg2009b}, were model-independent in the sense that they only relied on the presence of a physical momentum cut-off. The Bogoliubov parameter
was left as a free parameter and general bounds were derived on its magnitude for different inflationary Lagrangians. In this work we would like to 
specifically constrain the minimal cut-off modified vacuum state models for which we will need the expressions for the Bogoliubov parameters derived 
above.
 

\section{General bispectrum predictions}

As already referred to, in our previous work we calculated the results
for the three-point function under the assumption of an arbitrary
scale-invariant initial state modification in the NPH framework. The
presence of a physical high-energy momentum cut-off was assumed, which
identified an initial time for each comoving momentum mode where the
initial state was defined by introducing a $k$-independent, but
undetermined, Bogoliubov parameter. The presence of such a cut-off
alone already has major consequences for the three-point function, also known as the bispectrum in Fourier space. 
Whenever the Bogoliubov parameter is non-zero, particles are injected
at the initial (cut-off) time, which allows potential (self-) interactions to generate 
a large bispectrum amplitude at the time of horizon crossing. In the
calculations these effects show up as large enhancement factors that depend on powers
of  $\Lambda_c/H \gg 1$. Depending on the details of the inflationary
model under consideration the power of the enhancement factor could be as large as
three, resulting in a huge bispectrum amplitude. The only reason why
these large non-Gaussian signals have not yet been detected or ruled out is their 
specific (oscillatory) shape, which is orthogonal to any of the
non-Gaussian templates used in analyzing the CMB data so far. 
Proposals for better adapted templates and improved methods
to detect or constrain these oscillatory non-Gaussian signals have
recently been put forward \cite{Chen2010, Meerburg2010}, but have not
yet been applied to the available data, forcing us to concentrate on
the standard non-Gaussian shapes that have been constrained.
Projecting the large bispectrum onto any one of the available templates still 
allows one to derive reasonable constraints in general. 
In fact, in some cases the constraints are already stronger than those available from the power spectrum. 
Before discussing the results of the projections to the observational templates it should be pointed out that in principle the results 
only apply to the three-dimensional bispectrum. However, in general the changes resulting from the 
reduction to the two-dimensional sphere are minimal and can (partially) be
taken into account by introducing a weight function in the
three-dimensional analysis \cite{Fergusson:2008ra}.

For models with a small speed of sound ($c_s \ll 1$) the projections
to the local and orthogonal templates \cite{SSZ2009}, including the phase, were 
computed analytically \cite{Meerburg2009b}. To derive these results
the assumption was made that the modifications to the BD state were
strictly oriented in the $k_1$ direction, introducing a constant
$k_1\eta_0$ parameter, breaking the symmetry between the
different momenta in the triangle. Properly maintaining the symmetry
would instead require introducing three constant parameters $k_i\eta_0$, where $k_i$ is the
direction in which one has perturbed the BD vacuum state via a
Bogolyubov transformation. Consequently, one can factor out the
constant $k_i\eta_0$ for each perturbed state $\beta_{k_i}$. As
expected, the results of this more complete analysis that preserves
the symmetry between triangle momenta differ only slightly. Moreover,
we have ameliorated our (numerical) integration
methods and used the improved definition of the inner product
between shapes as proposed by \cite{Fergusson:2008ra}, which introduces a
weight function to simulate projection effects onto the 2D CMB sky. 
This increases the correlation between smooth and oscillating shapes
as oscillations are slightly damped by the weight function.  
 
General single field models are defined by a single (Lagrangian) function 
$P(X, \phi)$, where $X=g^{\mu\nu} \partial_\mu \phi \partial_\nu
\phi$. From this function $P(X, \phi)$ expressions can be derived for
the slow-roll parameters, the speed of sound and two other variables
$\Sigma$ and $\lambda$, whose ratio $\frac{2\lambda}{\Sigma}$ will 
appear in the three-point function \cite{Chen2007,singlefield_NG}. Introducing a modified NPH vacuum 
state the final enhancement factor in the projections turns out to be at 
best quadratic in $p \equiv k c_s \eta_0 = \Lambda_c c_s /H \gg
1$. The results, for an undetermined Bogoliubov parameter $\beta= | \beta | e^{i\delta}$, in the 
limit that $\frac{1}{c_s^2} \gg \frac{2\lambda}{\Sigma}$ and $p\gg1$ is  
\begin{eqnarray}
\Delta f^{\mathrm{local}}_{NL}&\simeq& 4\times 10^{-3} \, \frac{p^2|\beta|}{c_s^2} \, \cos(\delta),
\label{eq:constraint1}\\
\Delta f^{\mathrm{ort}}_{NL}&\simeq& - 6\times10^{-2} \, \frac{p^2|\beta|}{c_s^2}
\, \cos(\delta) \, .
\label{eq:constraint2}
\end{eqnarray} 
The additional terms that have been neglected are subleading in $p$ and are out of phase in the
sense that the phase parameter $\delta$ is shifted with $\pi/2$.  
Also note that these results do not reduce to the standard slow-roll results in the $c_s = 1$ limit.
Subdominant terms that have been neglected in the $c_s \ll 1$ limit
will turn in to the dominant contributions to the bispectrum in the $c_s=1$ limit, 
explaining why the $c_s=1$ limit of the above result does not
reproduce the slow-roll bispectrum. 
It is also important to stress that the DBI models are a special class
of small sound speed models for which $2\lambda /\Sigma = 1-1/c_s^2$
and as a result the leading contributions to the bispectrum cancel
exactly \cite{singlefield_NG}. The remaining contributions to the projections of the DBI
modified NPH initial state bispectrum are only linearly enhanced in the small 
$c_s$ limit and the details of the projections in the
DBI case can only be evaluated semi-analytically
\cite{Meerburg2009b}. This leads to DBI constraints that are far less
interesting as compared to general small sound speed models. We will
therefore concentrate on the general models and not discuss the
DBI results in detail. 

In the case of standard canonical single-field slow-roll inflation with a
specific dimension $8$ higher-derivative term \cite{Creminelli:2003iq} the enhancement factor appearing in
the bispectrum amplitude was found to be quadratic in $\Lambda_c/H$
\cite{Meerburg2009a}. However, in the projection to the available
observational templates $f^O_{NL} = \Delta(O,T) \, f^T_{NL}$, where
$\Delta(T,O)$ is the projection factor, one looses one power 
of the enhancement, leaving a linear enhancement in $\Lambda_c/H$ as far as the
projections are concerned. In addition the amplitude of the
enhancement is further reduced by slow-roll (via $\epsilon$) and the 
dimensionful coupling of the higher-derivative term. In these results
the effects of a (cut-off dependent) phase in the Bogoliubov parameter was neglected and for our
purposes here we would like to remedy this situation. To incorporate the phase 
we had to rely on numerical methods to determine the
projection factors. For the projection onto local non-Gaussianities we
obtain (in the limit of large $p\equiv \Lambda_c/H$)
\begin{eqnarray}
\Delta f^{\mathrm{local}}_{NL}&\simeq& \frac{5}{6} \, p \, \lambda \,
\epsilon \, |\beta| \, \left(\frac{M_{pl}^2 H^2}{\Lambda^4}\right) \cos(\delta),
\label{eq:constraint3}
\end{eqnarray} 
which, not surprisingly, is very similar to equation ($6.11$) in
\cite{Meerburg2009a}, except for the appearance of a phase dependence.
 
It is important to realize that the inclusion of the perturbative effect
of the dimension $8$ higher-derivitive operator is only consistent when
$\dot{\phi}/ \Lambda_c^2\lesssim 1$ \cite{Creminelli:2003iq}, which
translates into $\epsilon\geq \Lambda_c^4/ M_{pl}^2 H^2$.  The
power spectrum observations tell us that $\epsilon =
\frac{10^10}{8\pi^2}\frac{H^2}{M_{pl}^2}$, which implies that
$H/\Lambda_c < 10^{-2}$ to ensure a consistent higher-derivative expansion.
As a consequence the non-Gaussianities due to the presence of the
dimension $8$ higher-derivative term reach a maximal magnitude for 
$H/\Lambda_c \sim 10^{-2}$, yielding an equilateral amplitude of 
at most $f_{NL}^{\mathrm{equil}}\sim \mathcal{O}(1)$. On the other hand, 
modifying the initial state leads to an additional enhancement factor
that is proportional to the ratio between the UV cutoff $\Lambda_c$ and the scale of
inflation $H$. The required consistency of the higher-derivative
expansion, being inversely proportional to the cut-off scale
$\Lambda_c$, reduces this enhancement and as a consequence weakens 
the constraints on $\beta$. In fact, as we will see in the next section, for the 
minimal cut-off vacuum state proposal assumed in this paper the
constraint becomes practically meaningless. This should not come as much of a surprise, since in order to
boost non-Gaussianities from higher-derivative operators the cut-off scale needs to
be relatively close to the inflationary scale, whereas the non-Gaussian
amplitude as a consequence of initial state modifications grows as the
the hierarchy between the cut-off and the Hubble scale increases.   
For completeness, let us remark that the correlation between the orthogonal template and
the standard single-field higher-derivative bispectrum, being very similar, yields comparable constraints.

After this very short review and slight improvement of what is
known about bispectrum projections due to a generic NPH initial state, 
let us now derive the best available constraints on the parameters of 
minimal cut-off initial states.

\section{Bispectrum constraints on minimal cut-off vacuum states}

Having summarized the results for minimal cut-off initial states and the general projected non-Gaussian amplitudes 
due to arbitrary cut-off (or NPH) modified vacua, we would now like to
apply the most recent observational bounds to derive constraints on the 
parameters of the minimal cut-off initial state proposal.

To proceed we will make use of the following recent (2 $\sigma$) constraints on 
$f_{\mathrm{NL}}^{\rm local}$ and $f_{\mathrm{NL}}^{\rm ort}$ \cite{Smith:2009jr,Komatsu:2010fb} 
\begin{eqnarray*}
-10 & \leq f_{\mathrm{NL}}^{\rm local}\leq & 74,\\
-410 & \leq f_{\mathrm{NL}}^{\rm ort}\leq & 6
\end{eqnarray*}
The observational constraints on the equilateral amplitude
$f_{\mathrm{NL}}^{\rm equil}$ can be ignored, because the projection to the equilateral template 
turns out to be orders of magnitude smaller. One should keep in mind
that the power spectrum constraint on $|\beta|$ is roughly $10^{-2}$, 
which is based on the lack of evidence for oscillatory behavior with a
larger amplitude in the power spectrum \cite{Pspectrum-constraints}. 

\begin{figure}[htbp] 
   \centering
   \includegraphics[scale=0.6]{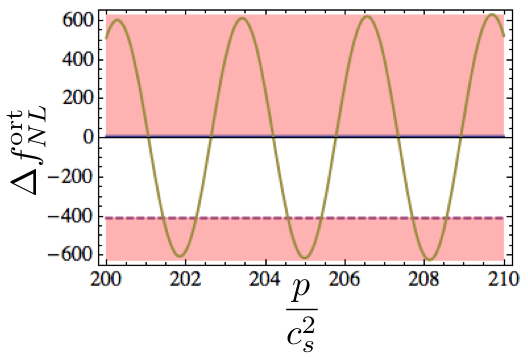} 
   \caption{Contribution of initial state modifications to orthogonal
     non-Gaussianities for various values of $p/c_s^2$.The red upper
     and lower parts in the diagram correspond to the observationally
     excluded regions.}
   \label{fig:small-cs}
\end{figure}

A notable feature of the results for the projections is that they
oscillate as a function of the phase $\delta$ of the Bogoliubov parameter 
(\ref{eq:constraint1}--\ref{eq:constraint2}). Different values 
for the phase can therefore result in constraints that deviate considerably. 
Indeed, for special values of the phase the projection vanishes and
the constraints disappear 
altogether\footnote{To be more precise, for those values one can no longer neglect
  the subleading terms in $p$, which are in effect $\pi/2$ out of phase and
  therefore maximize exactly when the leading terms
  minimize. Nevertheless, the fact that they are subleading in $p$
  implies that the constraints for those values of $p$ become
  practically meaningless.}. 
In minimal cut-off models the phase of the Bogoliubov parameter is
actually a function of the cut-off scale $\Lambda_c$ 
(\ref{eq:bogslowroll}--\ref{eq:bogsmallcs}) and as a consequence for
special values of $\Lambda_c$ the projected amplitudes would be 
significantly reduced, as can be seen in figure (\ref{fig:small-cs}). 
Since the precise value of the cut-off $\Lambda_c$ is unknown, the
constraints will be derived under the reasonable assumption  
that a typical value of $\Lambda_c$ is expected to be closer to an
(absolute value) maximum in the projection than to the special point 
where the projection vanishes. To implement this we will derive the
constraints based on an expression that is related to the absolute value
average over a single oscillatory domain in the parameter $\Lambda_c$.  
 
Let us first analyze general single field models with a small speed of
sound. The minimal cut-off vacuum state proposal predicts a Bogoliubov parameter with 
an absolute value approximately equal to $|\beta_k| \approx \frac{H}{2
  c_s \Lambda_c}$ and a phase $\delta = {i (\frac{3}{2}\pi - \frac{2
    c_s \Lambda}{H})}$. In the limit $p\equiv \Lambda_c \, c_s /H
\gg1$ we derive, to leading order in $p$,
\begin{eqnarray}
\Delta f^{\mathrm{local}}_{NL}&\simeq&- \, 2 \times10^{-3}\,
\frac{p}{c_s^2} \, \sin(2p),
\label{eq:constraint1v2}\\
\Delta f^{\mathrm{ort}}_{NL}&\simeq& 3 \times 10^{-2}\,
\frac{p}{c_s^2} \, \sin(2p),
\label{eq:constraint2v2}
\end{eqnarray}
The absolute value of the Bogoliubov coefficient has removed one
enhancement factor, resulting in a linear $p$ dependence of the
prefactor in front of the oscillatory $\sin(2p)$ term. As an example, the maximum contribution to
the orthogonal non-Gaussian template from initial state modifications has been
plotted in figure \ref{fig:small-cs}. As already alluded to, the
oscillations in the projection factor force us to work with the average over a single oscillatory 
domain in order to derive a constraint on the ratio $\Lambda_c /H$. 
Simply replacing the $\sin(2p)$ with the average $2/\pi$ the derived
bounds would only apply to about $60$ percent of the $p$ domain. If instead 
demanding that at least $90$ percent of the domain is included within
the constraints, we need to replace $\sin(2p)$ with $\sim 0.16$.  

The ambiguous sign due to the oscillatory behavior also prompts
us to consider the bound on the (largest) absolute value of the non-Gaussian 
amplitude, which for the local amplitude is $|f^{\mathrm{local}}_{NL}|
\leq 74$ and for the orthogonal type $|f^{\mathrm{ort}}_{NL}| \leq 410$.
Using these largest absolute values for the local and orthogonal amplitude and insisting on a $90$ percent coverage,  
an upper bound on the parameter $\Lambda_c \over H$ can be derived from local and orthogonal type non-Gaussianities
\begin{equation}
\mathrm{local}:\;\;{\Lambda_c \over H} \leq 2.5 \times 10^5 \, c_s
\;\; , 
\;\;\mathrm{orthogonal}:\;\;{\Lambda_c \over H} \leq 8.5  \times 10^4 \, c_s
\label{smallcsresults}
\end{equation} 
Clearly the strongest result is obtained from the orthogonal bound,   
becoming stronger as $c_s$ is smaller. In fact, the speed of sound
$c_s$ cannot be much smaller than $10^{-2}$ because that would produce a 
leading order equilateral non-Gaussian result in conflict with
observations. Applying that minimal value of $c_s$ the upper bound on 
$\Lambda_c \over H$ roughly equals 
\begin{equation}
{\Lambda_c \over H} \leq 10^3 \, ,
\end{equation}
Again we would like to stress that this is an {\it upper}
bound. The cut-off scale should be close to the inflationary Hubble scale to 
avoid too large an enhancement of the (projected) non-Gaussian signal,
with the exception of special higher values of $\Lambda_c$ for which the
phase $\delta$ is close to an integer number times $\pi$,
corresponding to $10$ percent of the parameter domain. 
This constraint is weaker than initially anticipated in
\cite{Meerburg2009b}, mainly because the phase in this case is
a function of $\Lambda_c$, producing oscillations in the
non-Gaussian signal, forcing us to introduce a factor related to the
average of the absolute non-Gaussian signal and the most conservative
(upper) bound on the absolute value of the observed constraints.

Incorporating the, admittedly crude, {\it lower} bound on ${\Lambda_c
  \, c_s\over H} > 10^2$ from analysis of the power spectrum, for a
general speed of sound one arrives at 
\begin{equation}
{10^2 \over c_s} \leq {\Lambda_c \over H} \leq 8.5  \times 10^4 \, c_s
\label{smallcsbounds}
\end{equation} 
There is a small window remaining for $c_s \sim 10^{-1}$, but for
small speed of sound models with $c_s \sim 10^{-2}$ one can
conclude that minimal initial state modifications are practically ruled out, except for those special values of the
phase at which the contribution to the orthogonal non-Gaussian template
(nearly) vanishes, corresponding to about $10$ percent of the
parameter domain.

To briefly illustrate the impact of the oscillatory behavior in this
analysis, consider neglecting the oscillating behavior, choosing $\sin
2p = 1$ for local non-Gaussianities and $\sin 2p = -1$ for orthogonal
non-Gaussianities. Again this would yield a strongest {\it upper}
bound from the orthogonal constraints that is roughly equal to
$\Lambda_c / H \leq 2 \times 10^2 \, c_s$, which for $c_s \sim
10^{-2}$ can essentially be ruled out from the non-Gaussian data alone and would 
in general be inconsistent with the power spectrum data that shows no
evidence of an oscillatory effect with an amplitude that large.

As already commented on, for DBI models the constraints weaken considerably, because the
enhancement of the non-Gaussian signal is reduced by a factor of
$p$. This results in no enhancement at all in the projection to the local (or
orthogonal) template, removing all dependence on the cut-off scale. 
As a a consequence the observational constraints do not result 
in (interesting) bounds on the cut-off scale $\Lambda_c$, which parameterizes
the initial state modification.  

Finally, let us briefly consider standard slow-roll inflation ($c_s=1$), including a dimension $8$ higher derivative term that, together with 
the initial state modification, is responsible for a large $p^2$
enhanced non-Gaussian signal. Inserting $|\beta|$ and the phase 
into eq. \eqref{eq:constraint2} one obtains
\begin{eqnarray}
f^{\mathrm{local}}_{NL}&\simeq& -\frac{5}{12} \sin( 2p )
\end{eqnarray}
All enhancement in $p\equiv\Lambda_c/H$ is lost and one is left with a
maximum contribution to the local non-Gaussian template that is of
(less than) order $1$, which is significantly below set constraints
on local type non-Gaussianities. As a consequence minimal initial state modifications in this
particular canonical single field inflationary model are not constrained
by the local non-Gaussian projection. In order to become sensitive to
the intrinsically large non-Gaussian signal set by the ratio $\Lambda_c/H$, templates better adapted to the 
theoretical non-Gaussian shape should be designed and compared to
the available data. 

\section{Conclusions}

To summarize, we have applied the currently best available constraints on local, orthogonal and equilateral
shape non-Gaussianities to derive bounds on the cut-off scale
parameterizing a (slightly generalized) minimal 
model of initial state modifications. The results strongly depend on the model of inflation
under consideration. Interesting bounds can be derived in the
context of general, non-DBI, single field small speed of sound models of
inflation. In that case, non-Gaussian constraints from the orthogonal
template alone lead to the following {\it upper} bound on the cut-off scale
\begin{equation}
{\Lambda_c \over H} \leq 8.5  \times 10^4 \, c_s \, .
\label{NG-bound}
\end{equation}
Combined with results from the equilateral template and the power
spectrum leaves only a very small window ($c_s\sim 10^{-1}$,
$\Lambda_c /H \sim 10^3$), and minimal initial state
modifications in general single field models with a small speed of
sound can almost be ruled out. To be more 
precise, the above upper bound is valid for $90$ percent of the
cut-off parameter domain, excluding small isolated regions covering in
total $10$ percent of the parameter domain where the projection to the orthogonal
template nearly vanishes as a consequence of the oscillating
nature of the projection. 

For DBI and a canonical single field model with a dimension $8$ higher
derivative operator the analysis does not lead to interesting
 non-Gaussian constraints. The reason is the complete lack of sensitivity of the
available templates to the large non-Gaussian signal of minimal
initial state modifications. Consequently, more suitable templates
have to be developed to efficiently probe (minimal) initial state
modifications in these models, perhaps along the lines of
\cite{Chen2010, Meerburg2010}.  Work in this direction is ongoing and
we hope to report on more effective non-Gausian templates and observational strategies
in the near future.

An important property that we have emphasized is that (minimal) initial state
modifications lead to a non-Gaussian signal that increases as the
separation between the cut-off scale and the inflationary Hubble
parameter increases. As a consequence non-Gaussian
constraints can only give rise to an upper bound on the cut-off scale
parameterizing initial state modifications arising due to new physics. Together
with lower bounds on the cut-off scale from other considerations
(including the power spectrum data or generic effective field theory arguments) this
carries a strong potential to rule out these types of initial
state modifications in the future as the non-Gaussian analysis and
corresponding constraints improve. 

At the same time the interesting feature
of an upper bound on the cut-off scale poses somewhat of a theoretical
conundrum, since it is obviously in conflict with the idea of
decoupling. Perhaps this should be interpreted as signalling a fundamental
flaw of these type of initial state modificatitions that might be
responsible for inconsistencies in the perturbative expansion of the quantum field
theory, conceivably ruling out these vacuum states. We hope to come back to 
this important issue in future work.


\section{Acknowledgements}
We would like to thank Pier Stefano Corasaniti for very beneficial and fruitful
discussions. PDM and JPvdS were supported in part by a van Gogh
collaboration grant VGP 63-254 from the Netherlands Organisation for
Scientific Research (NWO). PDM is supported by the Netherlands Organization
for Scientific Research (NWO) through a NWO-toptalent grant
021.001.040. The research of JPvdS is financially supported by Foundation of
Fundamental Research on Matter (FOM) grant 06PR2510. 


\begin{thebibliography}{10}

\bibitem{Chen2007}
X.~Chen, M.~x.~Huang, S.~Kachru and G.~Shiu,
``Observational signatures and non-Gaussianities of general single field inflation,''
JCAP {\bf 0701}, 002 (2007)
[arXiv:hep-th/0605045].

\bibitem{HolTol:2007}
  R.~Holman and A.~J.~Tolley,
  ``Enhanced Non-Gaussianity from Excited Initial States,''
  JCAP {\bf 0805} (2008) 001
  [arXiv:0710.1302 [hep-th]].

\bibitem{Meerburg2009a}
P.~D.~Meerburg, J.~P.~van der Schaar and P.~S.~Corasaniti,
``Signatures of Initial State Modifications on Bispectrum Statistics,''
JCAP {\bf 0905}, 018 (2009)
[arXiv:0901.4044 [hep-th]].

\bibitem{Meerburg2009b}
  P.~D.~Meerburg, J.~P.~van der Schaar and M.~G.~Jackson,
  ``Bispectrum signatures of a modified vacuum in single field inflation with a
  small speed of sound,''
  JCAP {\bf 1002} (2010) 001  
  arXiv:0910.4986 [hep-th].

\bibitem{Pspectrum-constraints}
L.~Bergstrom and U.~H.~Danielsson,
  ``Can MAP and Planck map Planck physics?,''
  JHEP {\bf 0212} (2002) 038
  [arXiv:hep-th/0211006].\\
  J.~Martin and C.~Ringeval,
  ``Exploring the superimposed oscillations parameter space,''
  JCAP {\bf 0501} (2005) 007
  [arXiv:hep-ph/0405249].\\
R.~Easther, W.~H.~Kinney and H.~Peiris,
  ``Observing trans-Planckian signatures in the cosmic microwave  background,''
  JCAP {\bf 0505} (2005) 009
  [arXiv:astro-ph/0412613].\\
D.~N.~Spergel {\it et al.}  [WMAP Collaboration],
  ``Wilkinson Microwave Anisotropy Probe (WMAP) three year results:
  Implications for cosmology,''
  Astrophys.\ J.\ Suppl.\  {\bf 170} (2007) 377
  [arXiv:astro-ph/0603449].\\
C.~Pahud, M.~Kamionkowski and A.~R.~Liddle,
  ``Oscillations in the inflaton potential?,''
  Phys.\ Rev.\  D {\bf 79} (2009) 083503
  [arXiv:0807.0322 [astro-ph]].

\bibitem{EGKS}
  R.~Easther, B.~R.~Greene, W.~H.~Kinney and G.~Shiu,
  ``Inflation as a probe of short distance physics,''
  Phys.\ Rev.\  D {\bf 64} (2001) 103502
  [arXiv:hep-th/0104102].\\
  R.~Easther, B.~R.~Greene, W.~H.~Kinney and G.~Shiu,
  ``Imprints of short distance physics on inflationary cosmology,''
  Phys.\ Rev.\  D {\bf 67} (2003) 063508
  [arXiv:hep-th/0110226].

\bibitem{NPH}
U.~H.~Danielsson,
``A note on inflation and transplanckian physics,''
Phys.\ Rev.\  D {\bf 66} (2002) 023511
[arXiv:hep-th/0203198].

\bibitem{Kinney2007}
W.~H.~Kinney and K.~Tzirakis,
  ``Quantum modes in DBI inflation: exact solutions and constraints from vacuum
  selection,''
  Phys.\ Rev.\  D {\bf 77} (2008) 103517
  [arXiv:0712.2043 [astro-ph]].

\bibitem{Chen2010}
  X.~Chen,
  ``Folded Resonant Non-Gaussianity in General Single Field Inflation,''
  arXiv:1008.2485 [hep-th].
  
\bibitem{Meerburg2010}
  P.~D.~Meerburg,
  ``Oscillations in the Primordial Bispectrum: Mode Expansion,''
  Phys.\ Rev.\  D {\bf 82} (2010) 063517
   arXiv:1006.2771 [astro-ph.CO].

\bibitem{Fergusson:2008ra}
  J.~R.~Fergusson and E.~P.~S.~Shellard,
  ``The shape of primordial non-Gaussianity and the CMB bispectrum,''
  Phys.\ Rev.\  D {\bf 80}, 043510 (2009)
  [arXiv:0812.3413 [astro-ph]].

\bibitem{SSZ2009}
L.~Senatore, K.~M.~Smith and M.~Zaldarriaga,
``Non-Gaussianities in Single Field Inflation and their Optimal Limits from the WMAP 5-year Data,''
arXiv:0905.3746 [astro-ph]. 

\bibitem{singlefield_NG}
D.~Seery and J.~E.~Lidsey,
``Primordial non-gaussianities in single field inflation,''
JCAP {\bf 0506} (2005) 003
[arXiv:astro-ph/0503692].

\bibitem{Creminelli:2003iq}
  P.~Creminelli,
  ``On non-gaussianities in single-field inflation,''
  JCAP {\bf 0310}, 003 (2003)
  [arXiv:astro-ph/0306122].

\bibitem{Smith:2009jr}
 K.~M.~Smith, L.~Senatore and M.~Zaldarriaga,
 ``Optimal limits on $ f_{\mathrm{NL}}^{local}$ from WMAP 5-year data,''
 JCAP {\bf 0909} (2009) 006
 [arXiv:0901.2572 [astro-ph]].
 
 \bibitem{Komatsu:2010fb}
  E.~Komatsu {\it et al.},
  ``Seven-Year Wilkinson Microwave Anisotropy Probe (WMAP) Observations:
  Cosmological Interpretation,''
  arXiv:1001.4538 [astro-ph.CO].



\end{thebibliography}
\end{document}